\documentclass{article}
\usepackage{amsmath,graphicx,mlspconf}
\usepackage{xcolor}
\usepackage{multirow}
\usepackage{amssymb}
\pdfoutput=1

\copyrightnotice{979-8-3503-7225-0/24/\$31.00 {\copyright}2024 IEEE}
\toappear{2024 IEEE International Workshop on Machine Learning for Signal Processing, Sept.\ 22--25, 2024, London, UK}

\title{COMPOSER STYLE-SPECIFIC SYMBOLIC MUSIC GENERATION USING VECTOR QUANTIZED DISCRETE DIFFUSION MODELS}
\name{Jincheng Zhang, Gy\"orgy Fazekas, Charalampos Saitis}
\address{Centre for Digital Music, Queen Mary University of London, UK}

\begin{document}

\maketitle

\begin{abstract}

Emerging Denoising Diffusion Probabilistic Models (DDPM) have become increasingly utilised because of promising results they have achieved in diverse generative tasks with continuous data, such as image and sound synthesis. Nonetheless, the success of diffusion models has not been fully extended to discrete symbolic music. We propose to combine a vector quantized variational autoencoder (VQ-VAE) and discrete diffusion models for the generation of symbolic music with desired composer styles. The trained VQ-VAE can represent symbolic music as a sequence of indexes that correspond to specific entries in a learned codebook. Subsequently, a discrete diffusion model is used to model the VQ-VAE's discrete latent space. The diffusion model is trained to generate intermediate music sequences consisting of codebook indexes, which are then decoded to symbolic music using the VQ-VAE's decoder. The evaluation results demonstrate our model can generate symbolic music with target composer styles that meet the given conditions with a high accuracy of 72.36\%. Our code is available at https://github.com/jinchengzhanggg/VQVAE-Diffusion.

\end{abstract}
\section{Introduction}
\label{sec:intro}

The application of deep learning has become increasingly prevalent in the field of symbolic music generation. Nevertheless, using deep learning to produce music with a desired composer style remains challenging. Composer style can be considered alongside genre as an important characteristic of music. While recent years have witnessed efforts dedicated to genre \cite{panteli2017towards}, music composer style studies remain very few in the literature. While a composer has the capability to create music across various genres, the distinctive nature of each composer's style remains largely unique~\cite{mukherjee2022composeinstyle}. 
Producing music that embodies a specific composer's style can cater to the personalised needs and preferences of diverse users.
For example, this will allow people to listen to musical compositions of specific composers who resonate with them. Therefore, there is a need to investigate the use of deep learning for music generation controlled by composer styles. 

Diffusion models \cite{ho2020denoising} have recently emerged as powerful generative models for continuous data tasks. They can not only produce samples with state-of-the-art quality, but also offer advantages such as providing a more comprehensive coverage of the sample space. For example, diffusion models surpass Generative Adversarial Network (GANs) in terms of the quality of generated samples for image synthesis\cite{dhariwal2021diffusion}. However, diffusion models' applications are generally confined to continuous data such as images and audio because of their Langevin-inspired sampling mechanisms. Therefore, the remarkable success of diffusion models has not been broadly extended to the generation of discrete symbolic music.

While Gu et al. \cite{gu2022vector} used a diffusion model combined with vector quantization (VQ) for continuous image synthesis, we adapted a VQ-VAE \cite{van2017neural} using music domain knowledge for encoding symbolic music to a sequence of indexes corresponding to specific vectors in a learned VQ-VAE's codebook. Subsequently, a discrete diffusion model was used to model the VQ-VAE's discrete latent space by reversing a forward diffusion process that progressively corrupts the intermediate music token sequence via a fixed Markov chain. To the best of our knowledge, our proposed method is the ﬁrst attempt that uses vector quantized diffusion models for discrete symbolic music generation. Furthermore, the proposed model allows users to control the generated music's composer style. Besides, we showcase our VQ-VAE is an effective method for representing long pianorolls as sequences of musical elements from a codebook with reduced length prior to their input to deep learning models. 




\section{RELATED WORK}
\label{sec:format}

Although diffusion models have achieved significant success in various generative tasks with continuous data, their uses for discrete data remain restricted. Only a few prior works have investigated the potential of diffusion models for discrete symbolic music generation \cite{mittal2021symbolicdiffusion,zhang2023fast,li2023melodydiffusion}. Mittal et al. used continuous diffusion models for infilling and unconditional generation of symbolic music \cite{mittal2021symbolicdiffusion}. To enable the use of continuous diffusion models, the authors trained a diffusion model on discrete symbolic music's continuous embeddings produced by a trained MusicVAE \cite{roberts2018hierarchical}. The distinct differences are that our work employs a VQ-VAE to represent symbolic music using sequential discrete tokens and we use a discrete diffusion model to model the VQ-VAE's discrete latent space. Similarly to the majority of previous works, MelodyDiffusion presented in \cite{li2023melodydiffusion} also simplifies the symbolic music generation task by training the diffusion models only on melodies, which may not be sufficient for expressing the subtleties of music.

There have been very few studies investigating vector quantization for symbolic music generation, among which Han et al. \cite{han2022symbolic} used a VQ-VAE to encode symbolic music loop into discrete representations and then trained an autoregressive model on the music loops' discrete latent codes for music loop generation. During training, autoregressive models employ "teacher-forcing" \cite{esser2021imagebart} to supply the correct ground truth token for prediction at each step. However, during the inference stage, each step depends on the previously sampled tokens to predict the current token. Hence, once a token is predicted during inference, it cannot be revised, leading to the problem of cumulative prediction errors in autoregressive models. Our diffusion model combined with vector quantization avoids error accumulation because the diffusion model is trained to learn and update the density distribution of all tokens and then sample all tokens simultaneously from the new distribution.


The attention paid to composer-specific music is also quite limited \cite{liang2017automatic,mukherjee2022composeinstyle}. A recent work \cite{mukherjee2022composeinstyle} used GAN models for a composer style transfer task. Our method is the first use of vector quantized diffusion models to generate symbolic music while providing users with flexible controls to generate symbolic music with desired composer styles, contributing to the progress on the new research topic of music composer style.

\section{METHOD}
\label{sec:pagestyle}

\subsection{Dataset}
\label{ssec:dataset}
Our vector quantized diffusion model (VQ-Diffusion) was trained using 300 long compositions by Liszt, Chopin and Schubert from the MAESTRO dataset \cite{hawthorne2018enabling}. For each composer, we used 100 MIDI performances of his compositions. Subsequently, we transformed the MIDI files into pianorolls using a sampling frequency of 32 samples per second. By using our VQ-VAE, the pianorolls were encoded into sequences of music tokens for training the vector quantized diffusion model.


\subsection{Model}
\label{ssec:network}

Standard diffusion models consist of a forward process and a reverse process. In the forward diffusion process, Gaussian noise is gradually added to the input data $\mathbf{x}_{0}$ in $\boldsymbol{T}$ diffusion steps via the pre-defined variance schedule $\beta_{1}, \ldots, \beta_{T}$. This process yields a sequence of increasingly noisy samples $\mathbf{x}_{1}$, \ldots, $\mathbf{x}_{T}$ with the same dimensionality as the data $\mathbf{x}_{0}$ until $\mathbf{x}_{T}$ is approximately Gaussian noise:
\begin{equation}
q\left(\mathbf{x}_{t} \mid \mathbf{x}_{t-1}\right)=\mathcal{N}\left(\mathbf{x}_{t} ; \sqrt{1-\beta_{t}} \mathbf{x}_{t-1}, \beta_{t} \mathbf{I}\right)
\end{equation}
The posterior probability of the forward process $q\left(\mathbf{x}_{1: T} \mid \mathbf{x}_{0}\right)$ defined in Equation (2) contains no trainable parameters, which is different from the encoder process of VAEs.

\begin{equation}
q\left(\mathbf{x}_{1: T} \mid \mathbf{x}_{0}\right)=\prod_{t=1}^{T} q\left(\mathbf{x}_{t} \mid \mathbf{x}_{t-1}\right)
\end{equation}

\begin{figure*}
\centering
\includegraphics[width=0.74\textwidth]{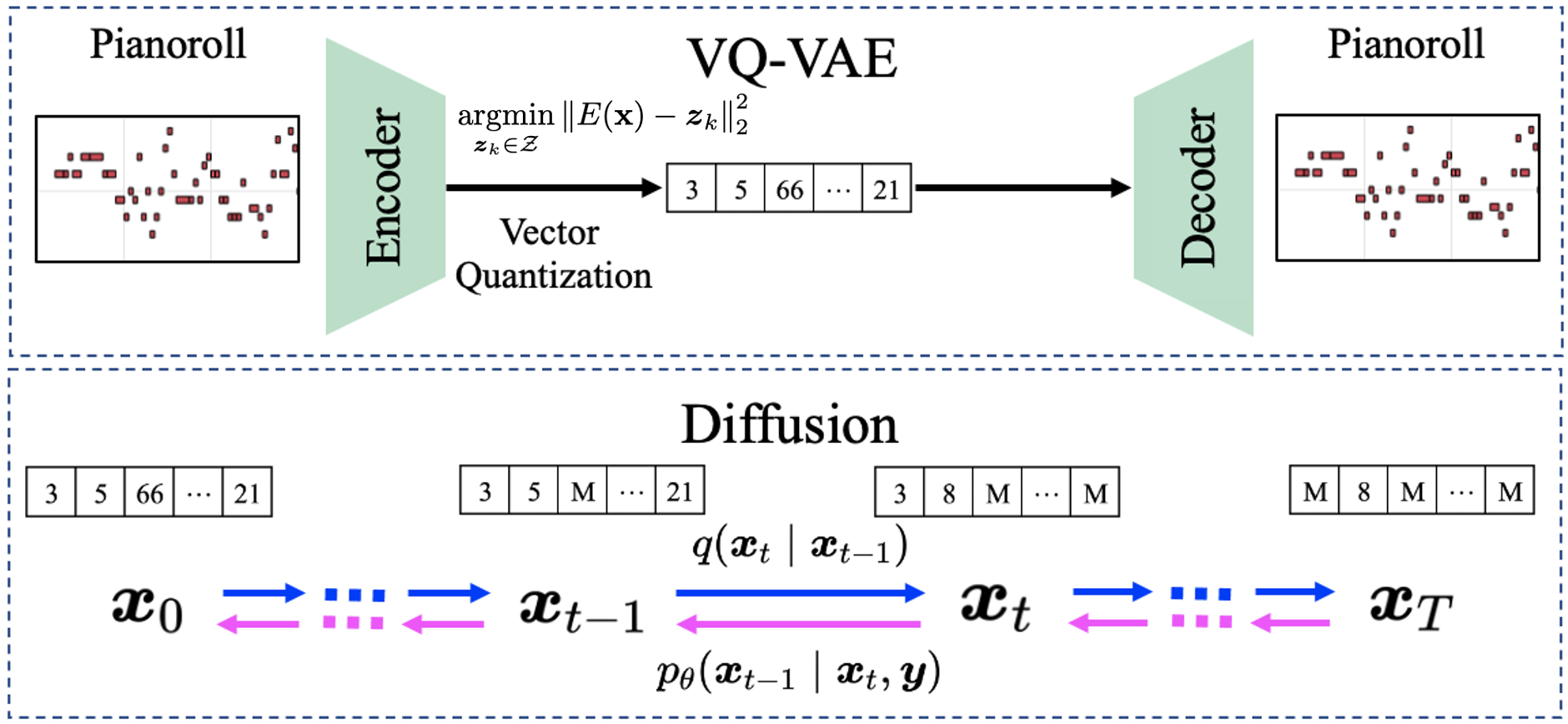}
\caption{\label{fig:vqdiff} Our approach uses a VQ-VAE to learn a codebook, whose composition is subsequently modeled with a discrete diffusion model. A Transformer is used as our denoising network.}
\end{figure*}


The reverse process uses a neural network such as a U-Net or a Tranformer to learn the conditioned probability distribution $p_{\theta}\left(\mathbf{x}_{t-1} \mid \mathbf{x}_{t}\right)=\mathcal{N}\left(\mathbf{x}_{t-1} ; \boldsymbol{\mu}_{\theta}\left(\mathbf{x}_{t}, t\right), \boldsymbol{\Sigma}_{\theta}\left(\mathbf{x}_{t}, t\right)\right)$. Then randomly sampled Gaussian noise $\mathbf{x}_{T} \sim \mathcal{N}(\mathbf{0}, \mathbf{I})$ can be iteratively denoised to generate target data: 
\begin{equation}
p_{\theta}\left(\mathbf{x}_{0: T}\right)=p\left(\mathbf{x}_{T}\right) \prod_{t=1}^{T} p_{\theta}\left(\mathbf{x}_{t-1} \mid \mathbf{x}_{t}\right)
\end{equation}
\begin{equation}
p_{\theta}\left(\mathbf{x}_{t-1} \mid \mathbf{x}_{t}\right)=\mathcal{N}\left(\mathbf{x}_{t-1} ; \mu_{\theta}\left(\mathbf{x}_{t}, t\right), \mathbf{\Sigma}_{\theta}\left(\mathbf{x}_{t}, t\right)\right)
\end{equation}

The distribution $p_{\theta}\left(\mathbf{x}_{t-1} \mid \mathbf{x}_{t}\right)$ can be learnt by minimising the negative log-likelihood of $p_\theta\left(\mathbf{x}_0\right)=\int p_\theta\left(\mathbf{x}_{0: T}\right) d \mathbf{x}_{1: T}$ via the variational lower bound (VLB):

\begin{equation} \label{eq:6}
\begin{split}
&\mathbb{E}_{q}[{D_{\mathrm{KL}}\left(q\left(\mathbf{x}_{T} \mid \mathbf{x}_{0}\right) \| p\left(\mathbf{x}_{T}\right)\right)}
{-\log p_{\theta}\left(\mathbf{x}_{0} \mid \mathbf{x}_{1}\right)}\\
+&\sum_{t>1} {D_{\mathrm{KL}}\left(q\left(\mathbf{x}_{t-1} \mid \mathbf{x}_{t}, \mathbf{x}_{0}\right) \| p_{\theta}\left(\mathbf{x}_{t-1} \mid \mathbf{x}_{t}\right)\right)}]
\end{split}
\end{equation}

We propose to combine VQ-VAE and discrete diffusion models for symbolic music generation, as shown in Fig. \ref{fig:vqdiff}. Compared to VAEs, a VQ-VAE includes an extra component named codebook $\mathcal{Z}=\left\{\boldsymbol{z}_k\right\}_{k=1}^K \in\mathbb{R}^{K \times d}$ containing a finite number of embedding vectors, where $K$ is the codebook size and $d$ is the code dimension. Given a symbolic music piece $\mathbf{x}$, it can be represented as a sequence of music tokens $\boldsymbol{z}_q$ using the encoding $\boldsymbol{z}=E(\mathbf{x})$ and the subsequent quantization $Q(\cdot)$ of each embedding $z$ to its closest codebook entry $z_k$. The encoded sequence length is usually much smaller than that of the raw input. 

\begin{equation}
\boldsymbol{z}_q=Q(\boldsymbol{z})=\underset{\boldsymbol{z}_k \in \mathcal{Z}}{\operatorname{argmin}}\left\|\boldsymbol{z}-\boldsymbol{z}_k\right\|_2^2
\end{equation}

Then the decoder $\tilde{\mathbf{x}}=G\left(\boldsymbol{z}_q\right)$ is used to reconstruct the symbolic music. The encoder, the decoder and the codebook are trained end-to-end using the loss function:

\begin{equation}
\begin{split}
\mathcal{L}_{\mathrm{VQVAE}}=\|\mathbf{x}-\tilde{\mathbf{x}}\|_1+\left\|\operatorname{sg}[E(\mathbf{x})]-z_q\right\|_2^2 \\
+\beta\left\|\operatorname{sg}\left[z_q\right]-E(\mathbf{x})\right\|_2^2
\end{split}
\end{equation}
where $\operatorname{sg}[\cdot]$ denotes the stop-gradient operation \cite{van2017neural}. VQ-VAE does not suffer from the issues of “posterior collapse”, often observed in many VAE models, which typically results from that the latents are ignored when the decoder is powerful \cite{van2017neural}.

The trained VQ-VAE can encode each pianoroll representation of input symbolic music into 1408 sequential embedding indexes from the codebook. We use discrete diffusion models to model the VQ-VAE's latent space. While the continuous diffusion models' forward process adds Gaussian noise to produce increasingly noisy data, we use a mask-and-replace diffusion strategy \cite{gu2022vector} for data corruption. Specifically, there is a probability of $\gamma_t$ that each ordinary token will be masked by $[MASK]$ token or it can probably be uniformly replaced by any one of the $K$ categories from the codebook with a chance of $K \beta_t$, leaving a probability of $\alpha_t=1-K \beta_t-\gamma_t$ that this token remains unchanged. Then the forward process of our discrete diffusion model can be described as $
q\left(\mathbf{x}_t \mid \mathbf{x}_{t-1}\right)=\boldsymbol{v}^{\top}\left(\mathbf{x}_t\right) \boldsymbol{Q}_t \boldsymbol{v}\left(\mathbf{x}_{t-1}\right)
$. $\boldsymbol{v}\left(\mathbf{x}\right)$ is a one-hot column vector where only the entry $\mathbf{x}$ is 1. The categorical distribution over $\mathbf{x}_t$ is given by $\boldsymbol{Q}_t \boldsymbol{v}\left(\mathbf{x}_{t-1}\right)$. $\boldsymbol{Q}_t$ is the transition matrix that defines the probabilities of transitioning from $\mathbf{x}_{t-1}$ to $\mathbf{x}_t$:

\begin{equation}
\boldsymbol{Q}_t=\left[\begin{array}{ccccc}
\alpha_t+\beta_t & \beta_t & \beta_t & \cdots & 0 \\
\beta_t & \alpha_t+\beta_t & \beta_t & \cdots & 0 \\
\beta_t & \beta_t & \alpha_t+\beta_t & \cdots & 0 \\
\vdots & \vdots & \vdots & \ddots & \vdots \\
\gamma_t & \gamma_t & \gamma_t & \cdots & 1
\end{array}\right]
\end{equation}

For the reverse process, a Transformer is used as our denoising network to estimate the distribution
$p_\theta\left(\tilde{\mathbf{x}}_0 \mid \mathbf{x}_t, \mathbf{y}\right)$, where $\mathbf{y}$ denotes the composer style.
We feed the composer style condition into the denoising network through the Adaptive Layer Normalization(AdaLN) \cite{ba2016layer} operator. 

All music tokens are either masked or random at the beginning of the inference stage. Based on the given composer style condition, the denoising diffusion process progressively estimates the probability density of music tokens step-by-step. At each step, the denoising network uses the contextual information of all tokens of the entire music predicted at the previous step to estimate a new probability density distribution. This updated distribution is then used to predict the tokens at the current step.





\subsection{Experiment Setup}
\label{ssec:experiment}
In our experiments, the codebook size $K$ of the VQ-VAE is 128. Our denoising network contains 16 transformer blocks with the dimension of 512. Each block contains a full attention layer and a feed forward network (FFN). We set the denoising timesteps $T$ = 100. Our network is optimized using AdamW \cite{loshchilov2017decoupled} with $\beta_1$ = 0.9 and $\beta_2$ = 0.96. The learning rate is set to 0.00045 after 5000 iterations of warmup, and a batch size of 16 is used.
 
\begin{figure}[b!]
\centering
\includegraphics[width=0.4\textwidth]{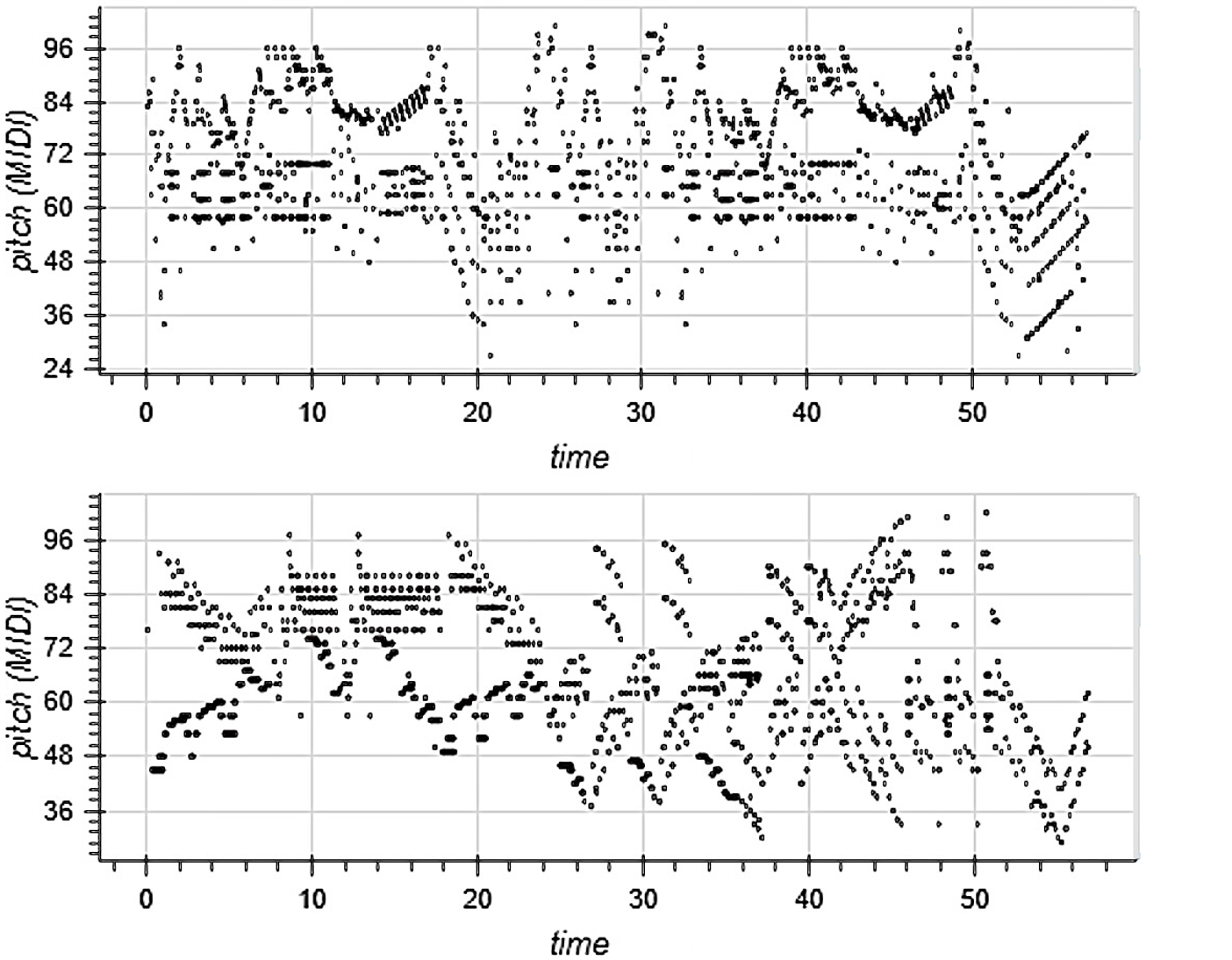}
\caption{\label{fig:music samples}Pianorolls generated by our vector quantized diffusion model (top) and samples from the training set (bottom). }
\end{figure}

\subsection{Objective Evaluation}

Music classifiers such as genre classifiers \cite{brunner2018symbolic} and composer classifiers \cite{mukherjee2022composeinstyle} have been widely used for evaluating the music style transfer systems to determine the style of music. This inspires us to perform an objective evaluation by training a one-dimensional convolutional neural network to identify the composer style of the generated music. The model consists of  five convolutional layers followed by one dense layer. Each convolution layer is followed by a ReLU activation and a batch normalization layer.

In addition, we adapted the Overlapping Area (OA) metric proposed by Choi et al.~\cite{choi2020encoding} to examine the similarity of feature distributions between the generations and the training dataset. We considered six features including note density (ND), pitch range (PR), mean pitch (MP), variation of pitch (VP), mean duration (MD) and variation of duration (VD). We evaluated the OAs using the estimated Gaussian probability density functions (pdfs) of the training dataset and our generated pieces, instead of comparing the estimated Gaussian pdfs of individual performances. The statistics of a feature distribution for the training dataset $O$ and the generations $G$ are defined as $\mathcal{N}(\mu_o, \delta_o^2)$ and $\mathcal{N}(\mu_g, \delta_g^2)$, respectively. The OA of the training dataset and generations for each feature can be computed using the following equation:
\begin{equation}
    OA(O, G) = 1 - \text{erf}(\frac{c-\mu_o}{\sqrt{2}\delta_o^2}) +\text{erf}(\frac{c-\mu_g}{\sqrt{2}\delta_g^2})
\end{equation}
where ${erf}$ is the Gauss error function and $c$ is the point of intersection between the two pdfs (assuming without loss of generality that $\mu_o > \mu_g$). The higher OA score reflects the higher similarity between two distributions.


\subsection{Listening Test}
\label{ssec:evaluation}

A listening test was conducted to further evaluate  the generated music. We used five metrics: Humanness assesses how closely the generated music resembles compositions by a human; Harmony refers to the way multiple notes are played together and how they create a collective and harmonious sound; Rhythm pertains to the pattern of sounds and silences in the music; Richness refers to the diversity and complexity of the music. We also included an Overall Preference metric.
30 participants (all adults) were recruited, including both musicians (30\%) and novice listeners without professional education or experience in music. 

For each of the three models, three generated symbolic music samples were rendered into audio for evaluation, following a similar setting with \cite{hung2021emopia}. Each participant was asked to listen to 12 music pieces, including nine generated samples and three pieces from the MAESTRO dataset. After listening to the generated pieces without knowing which model produced them, participants then rated the pieces on a 100-point scale in terms of the aforementioned subjective metrics. We included two same pieces to check the consistency of the participants. Ratings for the two repetitions are expected to be similar. Responses from participants who were not very consistent were excluded from further analysis.

\begin{figure}
\centering
\includegraphics[width=0.4\textwidth]{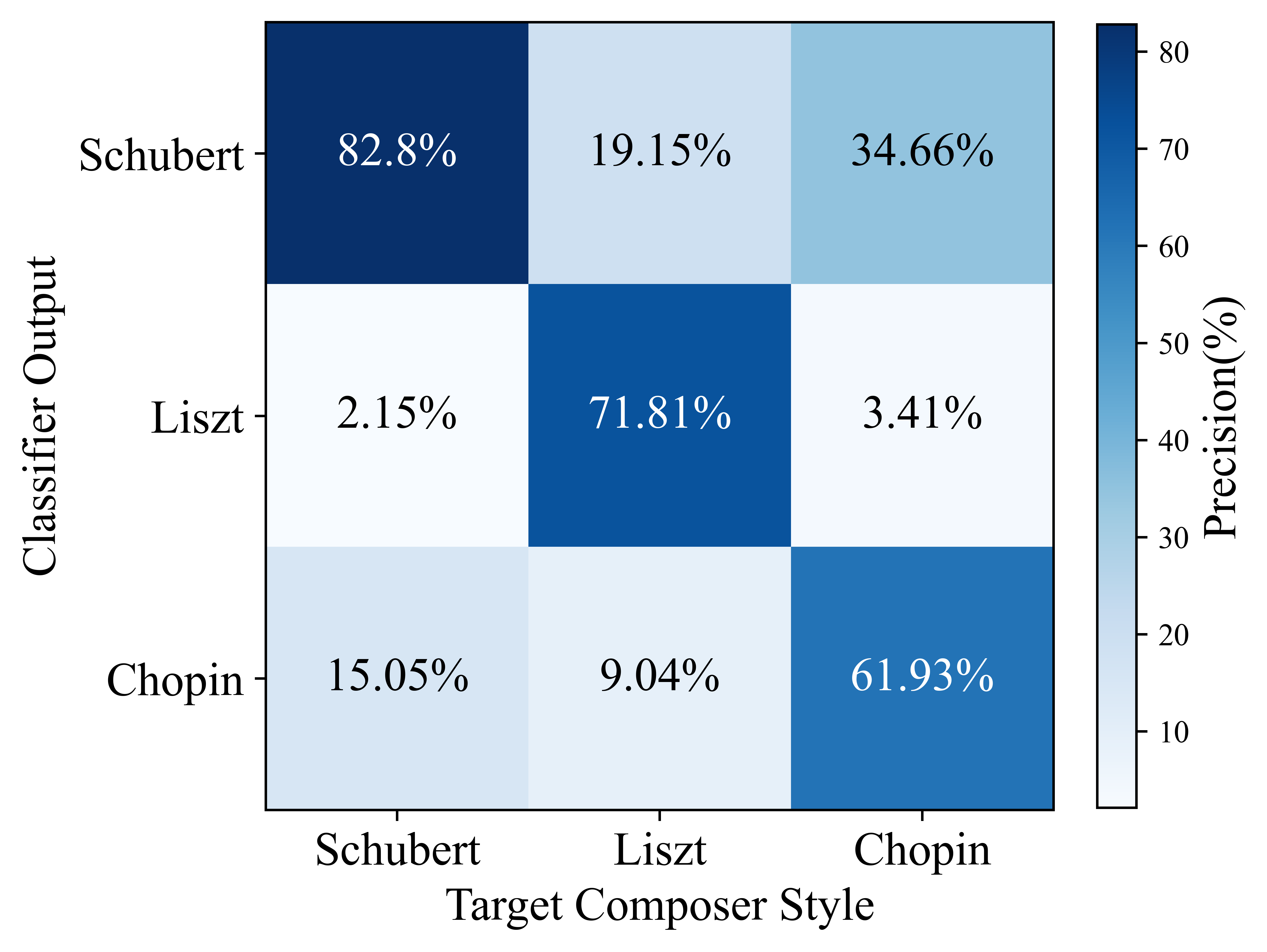}
\caption{\label{fig:matrice}
Accuracy calculated by assessing whether the generated pieces' composers predicted by the classifier meet the conditions fed to our vector quantized diffusion model.}
\end{figure}

\section{RESULTS \& DISCUSSION}
\label{sec:print}

         

\begin{table*}[t]
\centering
\resizebox{0.85\textwidth}{!}{%
\begin{tabular}{ccccccc}
\hline
\multirow{2}{*}{\textbf{Model}} & \textbf{Objective} & \multicolumn{5}{c}{\textbf{Subjective}} \\ \cline{2-7} 
 & Accuracy & Humanness & Harmony & Rhythm & Richness & Overall Preference \\ \hline
MAESTRO (real data) & -- & 72.82 $\pm$ 1.76 & 77.93 $\pm$ 1.43 & 73.71 $\pm$ 1.85 & 73.02 $\pm$ 2.27 & 76.73 $\pm$ 1.50\\ 
MusicTransformer & 60.67\% & 62.13 $\pm$ 1.98 & 61.07 $\pm$ 1.54 & 55.40 $\pm$ 1.78 & 59.67 $\pm$ 1.59 & 60.82 $\pm$ 1.34 \\ 
VQ-Transformer & 63.98\% & \textbf{65.18} $\pm$ \textbf{1.52} & 67.27 $\pm$ 1.53 & 62.78 $\pm$ 1.76 & 59.71 $\pm$ 2.14 & 64.42 $\pm$ 1.48 \\ 
Ours & \textbf{72.36\%} & 63.84 $\pm$ 1.84 & \textbf{69.58} $\pm$ \textbf{1.68} & \textbf{65.82} $\pm$ \textbf{1.82} & \textbf{72.16} $\pm$ \textbf{1.79} & \textbf{68.64} $\pm$ \textbf{1.55} \\ \hline
\end{tabular}%
}
\caption{Comparisons of different models for symbolic music generation in terms of objective and subjective metrics. The Accuracy metric is the average across the three composer styles. For the subjective metrics, the average is supplemented with standard error of the mean which indicates the accuracy of the average as an estimate of the population mean.}
\label{tab:my-table}
\end{table*}


We used our diffusion model to generate 200 samples from scratch for each composer style and then used our trained classifier to evaluate the composer style of the generated samples. The overall composer style control accuracy of our model is 72.36\%. This is a promising result since symbolic music generation controlled by composer style remains very challenging. Specifically, 82.80\% of the composer classifier’s outputs meet the condition fed to our generative model when the specified composer style is Schubert, as shown in Fig. \ref{fig:matrice}. 

We compared our VQ-Diffusion model with MusicTransformer \cite{huang2018music} which is the current state-of-the-art method for symbolic music generation. While the original MusicTransformer is unconditional, we achieved controllable music generation by simply concatenating music token embeddings with conditions. Regarding the ablation study, we replaced the discrete diffusion model in our VQ-Diffusion with a Transformer. This model is denoted as VQ-Transformer. All the three music generative models were trained on the same MAESTRO dataset for composer style-specific symbolic music generation. We also utilized the trained composer style classifier to assess how well the composer styles of the music pieces generated by the MusicTransformer and VQ-Transformer align with the conditions. Table 1 shows the evaluation results achieved by other models and ours. The classifier accuracy results suggest our proposed method achieves higher composer style accuracy than VQ-Transformer and MusicTransformer. This advantage is probably attributed to that diffusion model decomposes the controllable generation process into a sequential denoising diffusion steps where each denoising step is relatively simpler to model.

\begin{table}[b]
    \centering
    \resizebox{0.98\columnwidth}{!}{
    \begin{tabular}{c c c c c c c c}
        \hline
        \textbf{Model} & \textbf{ND} & \textbf{PR} & \textbf{MP} & \textbf{VP} & \textbf{MD} & \textbf{VD} & \textbf{Avg} \\ \hline
        MusicTransformer & 0.71 & 0.80 & 0.96 & 0.87 & 0.58 & 0.76 & 0.78\\ 
        VQ-Transformer & 0.83 & 0.80 & \textbf{0.97} & 0.81 & 0.88 & 0.87 & 0.86\\ 
        Ours & \textbf{0.89} & \textbf{0.91} & 0.94 & \textbf{0.88} & \textbf{0.93} & \textbf{0.95} & \textbf{0.92}\\ \hline
    \end{tabular}
    }
    \caption{Overlapping Area (OA) similarity comparing the distributions of training data and generations by the three different models.}
    \label{tab:OA comparision}
\end{table}


As the OA results presented in Table \ref{tab:OA comparision}, our model obtains higher OA similarity between the distributions of generated samples and the training set. This suggests our vector quantised diffusion model effectively learned the distribution of composition-related features from the training data. Overall, the composer style classifier and OA metric demonstrate our diffusion model can generate music with composer styles that are consistent with given conditions at a high level of accuracy, while achieving the state-of-the-art OA similarity of feature distributions.



\begin{figure}
\centering
\includegraphics[width=0.4\textwidth]{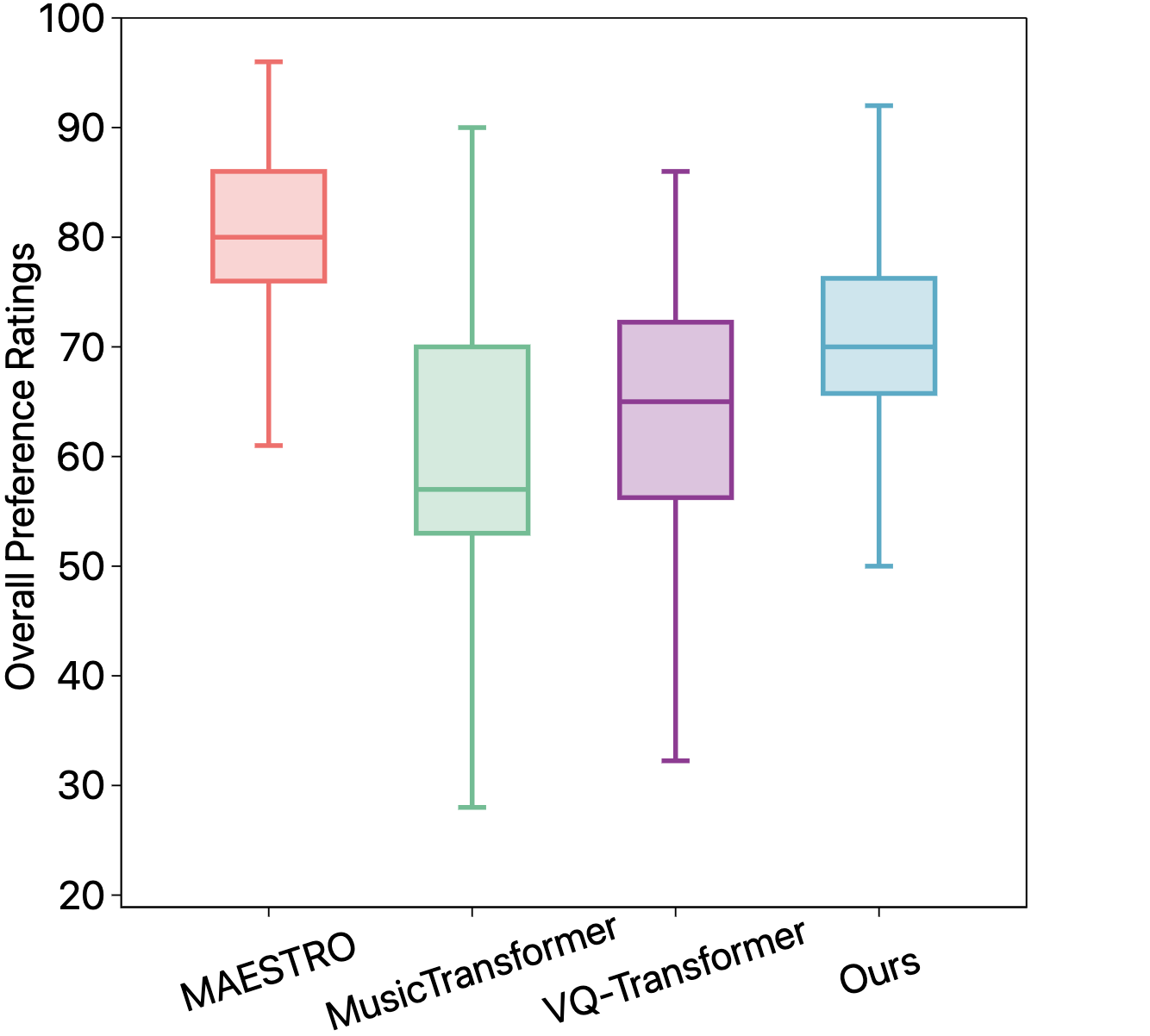}
\caption{\label{fig:boxplot}
Distribution of Overall Preference ratings for our vector quantized diffusion model and other methods.}
\end{figure}


The subjective listening test results are shown in Table 1. The music samples generated by the proposed VQ-Diffusion show Humanness comparable with the MusicTransformer and VQ-Transformer. Xiao et al. \cite{xiao2022tackling} demonstrated diffusion models' advantage of generating diverse samples. Our model outperforms the other two models with respect to Richness, indicating high diversity and complexity in the music it generates. This is also a good agreement with the objective OA metric as Table 2 shows VQ-Diffusion performs well in terms of variation of pitch and pitch range which probably relate to the perception of richness. Moreover, our VQ-Diffusion scores higher than MusicTransformer and is comparable with VQ-Transformer in terms of Harmony, Rhythm and Overall Preference. Fig. \ref{fig:boxplot} visualizes the rating distribution of the Overall Preference metric. It suggests that the proposed model has a higher median rating than both the VQ-Transformer and MusicTransformer.

\section{CONCLUSION AND FUTURE WORK}

In this paper, a novel generative music system is presented using discrete diffusion models and vector quantization. We model the VQ-VAE's latent space using discrete diffusion models, avoiding the autoregressive models' issue of error accumulation. This work can facilitate a wider adoption of diffusion models in discrete symbolic music generation. Moreover, our method allows the user to specify the generated music's composer style and achieves the state-of-the-art accuracy of composer style control. We leave symbolic music generation conditioned on more composer styles and text-to-music generation for future work. Besides, some conditioning approaches \cite{ho2022classifier} for diffusion models can also be explored.

\section{ACKNOWLEDGEMENTS}
We would like to thank Jingjing Tang for helping with the data preprocessing. The first author is supported by the China Scholarship Council.






\vfill
\pagebreak

\bibliographystyle{IEEEbib}
\bibliography{strings,refs}

\end{document}